\documentclass[a4paper]{article}
\usepackage[utf8]{inputenc}
\usepackage{graphicx}				
\usepackage{amsmath}
\usepackage[caption=false]{subfig}
\usepackage{cite}

\newcommand{\beq}{\begin{equation}}
\newcommand{\eeq}{\end{equation}}
\newcommand{\bea}{\begin{eqnarray}}
\newcommand{\eea}{\end{eqnarray}}
\newcommand{\beqa}{\begin{eqnarray}}
\newcommand{\eeqa}{\end{eqnarray}}

\newcommand{\pythia}{{\sc Pythia}}
\newcommand{\herwig}{{\sc Herwig}}
\newcommand{\cascade}{{\sc Cascade}}
\newcommand{\katie}{{\sc Ka\hspace{-0.2ex}Tie}}
\newcommand{\sherpa}{{\sc Sherpa}}

\newcommand{\kt}{$k_{\rm T}$}
\newcommand{\muF}{\mu_{\rm F}}
\newcommand{\KT}[1]{k_{\rm T#1}}
\newcommand{\PT}[1]{p_{\rm T#1}}

\newlength{\dinwidth}
\newlength{\dinmargin}
\usepackage{color}
\usepackage[makeroom]{cancel}
\usepackage[normalem]{ulem}
\definecolor{pkcolor}{rgb}{0,0.1,0.7}

\newcommand\pkout{\marginpar{\color{pkcolor}$\clubsuit$}\bgroup\markoverwith{\color{pkcolor}{\rule[04ex]{2pt}{0.8pt}}}\ULon}

\definecolor{kkcolor}{rgb}{1,0,0}

\newcommand\kkout{\marginpar{\color{kkcolor}$\clubsuit$}\bgroup\markoverwith{\color{kkcolor}{\rule[04ex]{2pt}{0.8pt}}}\ULon}

\definecolor{pvmcolor}{rgb}{0.7,0,0.7}

\definecolor{hvhcolor}{rgb}{0.0,0,0.7}

\title{ Trijets in $\KT{}$-factorisation: matrix elements \textsl{vs}\@ parton shower}
\author{
 H. Van Haevermaet$^a$, A. Van Hameren$^b$, P. Kotko$^c$,\\ K. Kutak$^b$, P. Van Mechelen$^a$ \\ \\
$^a${\it  University of Antwerp, Particle Physics group,}\\ 
    {\it Groenenborgerlaan 171, 2020 Antwerpen, Belgium} \\ \\
$^b$ {\it Institute of Nuclear Physics, Polish Academy of Sciences} \\
     {\it  Radzikowskiego 152, 31-342 Krak\'ow, Poland } \\ \\
$^c${\it AGH University Of Science and Technology, Physics Faculty,} \\ 
    {\it Mickiewicza 30, 30-059 Krak\'ow, Poland} 
}

\date{}

\begin{document}
\maketitle

\begin{abstract}
We study 3-jet event topologies in proton-proton collisions at a centre-of-mass energy of $\sqrt{s} = 13 {\rm\ TeV}$ in a configuration, where one jet is present in the central pseudorapidity region ($|\eta| < 2.0$) while two other jets are in a more forward (same hemisphere) area ($|\eta| > 2.0$).
We compare various parton level predictions using: collinear factorisation, $\KT{}$-factorisation with fully off-shell matrix elements and the hybrid framework. We study the influence of different parton distribution functions, initial state radiation, final state radiation, and hadronisation. We focus on differential cross sections as a function of azimuthal angle difference between the leading dijet system and the third jet, which is found to have excellent sensitivity to the physical effects under study.
\end{abstract}

\section{Introduction}
\label{sec:Intro}

Thanks to the hadron-parton duality, jet production processes at the Large Hadron Collider (LHC) are the best tools to study perturbative Quantum Chromodynamics (QCD) (for a review see \cite{Sapeta:2015gee}). The relation between experimental observables and the QCD degrees of freedom is, however, highly nontrivial: due to colour confinement, 
the partonic content of hadrons is unknown from first principles, while asymptotic freedom of quarks and gluons allows to study many aspects of hadronic physics perturbatively \cite{Collins:2011zzd}. So-called factorisation theorems make this relation formal and allow for a systematic approach. In the case of some of the  simplest  observables, like hadron structure functions or the cross section for inclusive production of very energetic jets, a suitable, well established formalism is provided by the so-called collinear factorisation theorem (for a review see \cite{Collins:2011zzd}).
Using it, the cross sections for sufficiently inclusive processes can be calculated in terms of collinear Parton Distribution Functions (PDFs) and perturbative on-shell amplitudes for the scattering of quarks and gluons. Less inclusive observables, or processes involving multiple large scales, however, require different formalisms utilising various all-order resummations of potentially large logarithms. At the LHC, many jet observables are subject to resummation and other corrections reaching beyond collinear factorisation (e.g.\@  multiple partonic interactions). Among other reasons, this is due to the overall very large centre-of-mass energy, as well as the ability to measure small jet transverse momenta, $\PT{}$, with good resolution. In addition, good jet reconstruction capabilities allow to measure the azimuthal angle between jets, which is sensitive to soft gluon emissions and to the transverse momentum of partons inside hadrons. In this paper we will focus on such observables, as a sensitive probe of parton dynamics. 

A formal theoretical framework dealing with parton transverse momenta, \kt, to leading power accuracy is the Transverse Momentum Dependent (TMD) factorisation theorem \cite{Collins:1984kg,Collins:2011zzd} (for recent applications see \cite{Scimemi:2019cmh,Bertone:2019nxa,Bacchetta:2019sam}), which however holds to all-orders only for processes with a total of at most two hadrons in the initial or final state. 
There are  less strict formalisms (working to leading logarithmic accuracy) like soft gluon resummation  or  \kt-factorisation (also called High Energy Factorisation (HEF)), \cite{Collins:1991ty,Catani:1990eg}. The latter is suitable for collisions with very large centre-of-mass energy and takes into account power corrections.  
On the phenomenology side,  general purpose Monte Carlo generators, like \pythia \cite{Sjostrand:2006za,Sjostrand:2014zea} , \herwig \cite{Bahr:2008pv,Bellm:2015jjp}, and \sherpa \cite{Bothmann:2019yzt} use collinear factorisation in combination with parton showers to generate partons with non-zero \kt.

This variety of approaches with different realisations of potentially similar mechanisms calls for a 
detailed comparison and validation, as well as confrontation with experimental data.
In this paper we investigate the \kt-factorisation approach, as well as collinear factorisation supplemented with parton showers, in the context of \emph{trijet production processes}. The case of dijet production was addressed in \cite{Bury:2017jxo}. In that paper, it has been studied to what extend calculations using unintegrated parton densities with off-shell matrix elements result in similar predictions as including higher order contributions in collinear calculations.  It turned out that including initial state TMD parton showers together with conventional final state parton showers gave
a remarkably good description of the measurements.
In the present  paper we ask different questions. Trijet events, being less inclusive than dijet events, are interesting to investigate the sensitivity to Sudakov resummation and to explore to what extend matrix elements with lower multiplicity supplemented with parton showers can mimic the predictions obtained with higher multiplicity matrix elements. As we shall show, the azimuthal angle distribution between the two leading jets and a third jet is very sensitive to the underlying models, having thus the discriminating power needed to address the questions above. 

This paper is organised as follows. In Section~\ref{sec:Framework} we review 
\kt-factorisation in the context of trijet production. In Section~\ref{sec:Setup} we describe the kinematic setup and Monte Carlo event generator programs used in our calculations.  Sections~\ref{sec:PartonLevel}-\ref{sec:Multiplicity} are devoted to a detailed study of the influence of various aspects of the calculations: parton-level, hadron-level and the multiplicity of the hard process. Finally, Section~\ref{sec:Summary} concludes with a summary.

\section{Theoretical framework}
\label{sec:Framework}
The \kt-factorisation formula applied to the case of inclusive trijet production at leading order reads:
\begin{multline}
\sigma_{{\rm pp\rightarrow 3jet+}X} =
\sum_{i,j} \int \frac{dx_1}{x_1}\,\frac{dx_2}{x_2}\, d^2 \KT{1}\, d^2 \KT{2}\,\,  \mathcal{F}_i(x_1,|\vec{k}_{\rm T1}|,\muF)\, \mathcal{F}_j(x_2,|\vec{k}_{\rm T2}|,\muF)  \\
\times \frac{1}{2 \hat{s}}\int \prod_{l=1}^3 \frac{d^3 k_l}{(2\pi)^3 2 E_l} \, \Theta_{\mathrm{3jet}}\left(\{k_l\}\right)\,
 \left| \overline{\mathcal{M}}(i^*,j^* \rightarrow \{k_l\})\right|^2 
\\ \,\times (2\pi)^4 \delta^{(4)}\left(\,\sum_{m=1}^2(x_m P_m+\KT{\,m}) - \sum_{l=1}^3 k_l \right) \, . 
\label{kt_cross}
\end{multline}

Here $\mathcal{F}_i(x,\KT{},\muF)$ is an unintegrated PDF (also called sometimes transverse momentum dependent PDF) for a type of parton $i$. 
Similarly as in collinear factorisation, it depends on the longitudinal fraction $x$ of the hadron momentum $P$ carried by the parton, but here a new degree of freedom appears -- the 
magnitude of the parton transverse momentum $\KT{}$, i.e.\@ the momentum perpendicular to the collision axis ($P\cdot \KT{} =0$). Originally, the unintegrated PDFs did not depend on the factorisation scale $\muF$ \cite{Fadin:1975cb,Balitsky:1978ic}, as they were applied to inclusive charm quark production \cite{Catani:1990eg}. However, if we want to apply this formalism to jets, where $\muF$ is of the order of the rather large average transverse momentum of jets $\PT{}$, we need to include an evolution in $\muF$. This is achieved by means of the Sudakov form factor which is the kernel of the DGLAP evolution. 
Its exact form used on the top of the  $\KT{}$-dependent gluon densities following ideas developed in \cite{Kimber:2001sc,vanHameren:2014ala,Kutak:2012rf} assumes the following form
 \begin{equation}
T_s(\muF^2,k_{\mathrm{T}}^2)=\exp\left(-\int_{k_{\mathrm{T}}^2}^{\muF^2}\frac{dk_{\mathrm{T}}^{\prime 2}}{k_{\mathrm{T}}^{\prime 2}}\frac{\alpha_s(k^{\prime 2})}
{2\pi}\sum_{a^\prime}\int_0^{1-\Delta}dz^{\prime}P_{a^\prime a}(z^\prime)\right)
\end{equation}
where $\Delta=\frac{\muF}{\muF+k_{\mathrm{T}}}$ and $P_{a^\prime a}$ is a splitting function with subscripts $a^\prime a$ specifying the type of transition. In the $gg$ channel one multiplies  $P_{gg}(z)$ by $z$ \cite{Kimber:2001sc,Watt:2003mx}.
In the equation above the $\muF$ introduces a hard scale dependence and is linked to the hard process.
Effectively, the above Sudakov form factor provides resummation of logs of $|k_{\mathrm{T}}|/\muF$. 
The next essential component of formula (\ref{kt_cross}) consists of the off-shell gauge invariant amplitudes $\mathcal{M}(i^*,j^*\rightarrow \{k_l\})$ for scattering of off-shell partons $i^*,j^*$ to produce a three-parton final state. The methods to calculate such processes in a gauge invariant way were developed in \cite{Antonov:2004hh,vanHameren:2012if,vanHameren:2012uj,Kotko:2014aba,vanHameren:2015bba}.
The $\Theta_{\mathrm{3jet}}$ function is the jet algorithm function that prevents entering singular regions of the phase space and provides kinematic cuts.

The factorisation formula for trijet case (\ref{kt_cross}) is valid when $x_1$ and $x_2$ are not too large and not too small (for in dijet case in this region see \cite{Deak:2010gk,Deak:2011ga,Kutak:2012rf,vanHameren:2014ala}) -- in the latter case, complications arise due to very large gluon densities leading to saturation and nonlinear evolution equations \cite{Gribov:1984tu,McLerran:1993ka,Kovchegov:1999yj,Balitsky:1995ub,Kovner:1999bj,Iancu:2000hn}.
Since our study is limited to central and mid rapidity for at least one parton, we avoid the saturation regime.
In our investigations we will also use the \emph{hybrid} HEF formalism \cite{Dumitru:2005gt,Deak:2009xt}. This framework is relevant when $x_1\gg x_2$, which allows to replace the unintegrated PDF for the large $x$ parton by the collinear one, formally, by integrating it over $\KT{}$. In this approach, trijet calculations have been done previously in \cite{vanHameren:2013fla}, albeit
only considering gluons as initial-state off-shell partons.

\section{Kinematics and Monte Carlo event generator setup}
\label{sec:Setup}

In this paper we will use the parton-level event generator \katie \cite{vanHameren:2016kkz} to obtain numerical values for hard scattering matrix elements. In case of on-shell kinematics the output is propagated to \pythia 8\ to add initial state radiation (ISR), final state radiation (FSR), multiple partonic interactions (MPI), and hadronisation effects. For the full off-shell matrix element configurations the output of \katie\ is propagated to \cascade 3 \cite{Jung:2010si} to add ISR, FSR, and hadronisation.

In all samples, the anti-$\KT{}$\  algorithm \cite{Cacciari:2008gp} with distance parameter $R=0.4$ is used to cluster particles into jets with  $\PT{} > 20 {\rm\ GeV}$ and pseudorapidity $|\eta| < 4.7$.  We further require to have one jet present in the central pseudorapidity region ($|\eta_{\rm{c}}| < 2.0$), and two other jets in a more forward area ($|\eta_{\rm{f1,f2}}| > 2.0$) with both in the same pseudorapidity hemisphere ($\eta_{\rm{f1}}\cdot\eta_{\rm{f2}} > 0$). Finally, the leading jet is required to have $\PT{} > 35$ GeV. 

Five processes are included in the $2 \rightarrow 3$ matrix element calculations: $gg \rightarrow ggg$, $gg \rightarrow gq\bar{q}$, $qg \rightarrow ggq$, $qg \rightarrow qq\bar{q}$, and $qg \rightarrow qq'\bar{q}'$, with $q$ and $q'$ representing quarks of a different flavour. 
These calculations are compared to predictions obtained by using $2 \rightarrow 2$ hard scattering processes complemented with parton showers to account for the third jet. In that case we consider the $gg \rightarrow gg$, $gg \rightarrow q\bar{q}$, and $qg \rightarrow qg$ subprocesses. The renormalisation and factorisation scales are set to ${\rm H}_{\rm T}/2$, with ${\rm H}_{\rm T}$ the scalar sum of all jet transverse momenta. Note that during the generation of the samples a lower $\PT{}$ threshold on the produced partons is used to allow for migration effects.

Various PDF sets are used: CT10NLO obtained from LHAPDF6 \cite{Buckley:2014ana}, and MRW-CT10NLO \cite{Watt:2003mx,Bury:2017jxo}\footnote{See the discussion on some subtleties of MRW type of unintegrated parton densities \cite{Hautmann:2019biw,Nefedov:2020ecb,Golec-Biernat:2018hqo,Guiot:2019vsm}.}
and PB-NLO-HERAI+II-2018-set2 \cite{Hautmann:2017fcj,Martinez:2018jxt} from TMDlib \cite{Hautmann:2014kza}. The latter unintegrated PDF enables us to study ISR effects in the HEF framework, as it can be used in \cascade 3\ to produce a full flavour unintegrated parton density based parton shower evolution.  While applying the hybrid framework  for the matrix element calculations, a linear and nonlinear version of the unintegrated Kutak-Sapeta (KS) PDFs is used \cite{Kutak:2012rf}. These PDFs however only contain gluon information, and can thus not be used to produce a full flavour parton shower evolution. We will therefore only include these PDFs during our parton level studies, in which the hybrid framework implies that the initial gluon is taken to have off-shell kinematics, while the other initial parton has on-shell kinematics and uses the collinear CT10NLO PDF. This will also allow us to estimate whether we can safely neglect nonlinear effects and continue with gluon densities obtained from linear evolution equations. An additional variant of the KS PDFs (called KShardscale-lin and KShardscale-nonlin) is available where also Sudakov resummation is taken into account \cite{Kutak:2014wga}.
As mentioned before, the Sudakov resummation is needed since there is an ordering in the hard scale $\muF$ and the imbalanced $\KT{}$ of initial state partons. 
The Sudakov form factor that we use is essentially valid in the region where $\muF$ is larger than the transverse momentum of the incoming gluon. The construction of the KShardscale unintegrated gluon density includes a $\theta$ function separating the two regions. The detailed formula can be found in \cite{vanHameren:2014ala}.
The formula for the resulting gluon density dependent on $\KT{}$, $x$, $\muF$ can be found in \cite{Kutak:2014wga}.
It has been recently observed that, even in the $\KT{}$-factorisation approach, the Sudakov form factor a gives rather large contribution to azimuthal angle related final state observables \cite{vanHameren:2014ala,vanHameren:2019ysa,Marquet:2019ltn}.

\section{Parton level predictions}
\label{sec:PartonLevel}

\begin{figure}[t!]                                   \centerline{\subfloat[]{\includegraphics[width=0.52\textwidth]{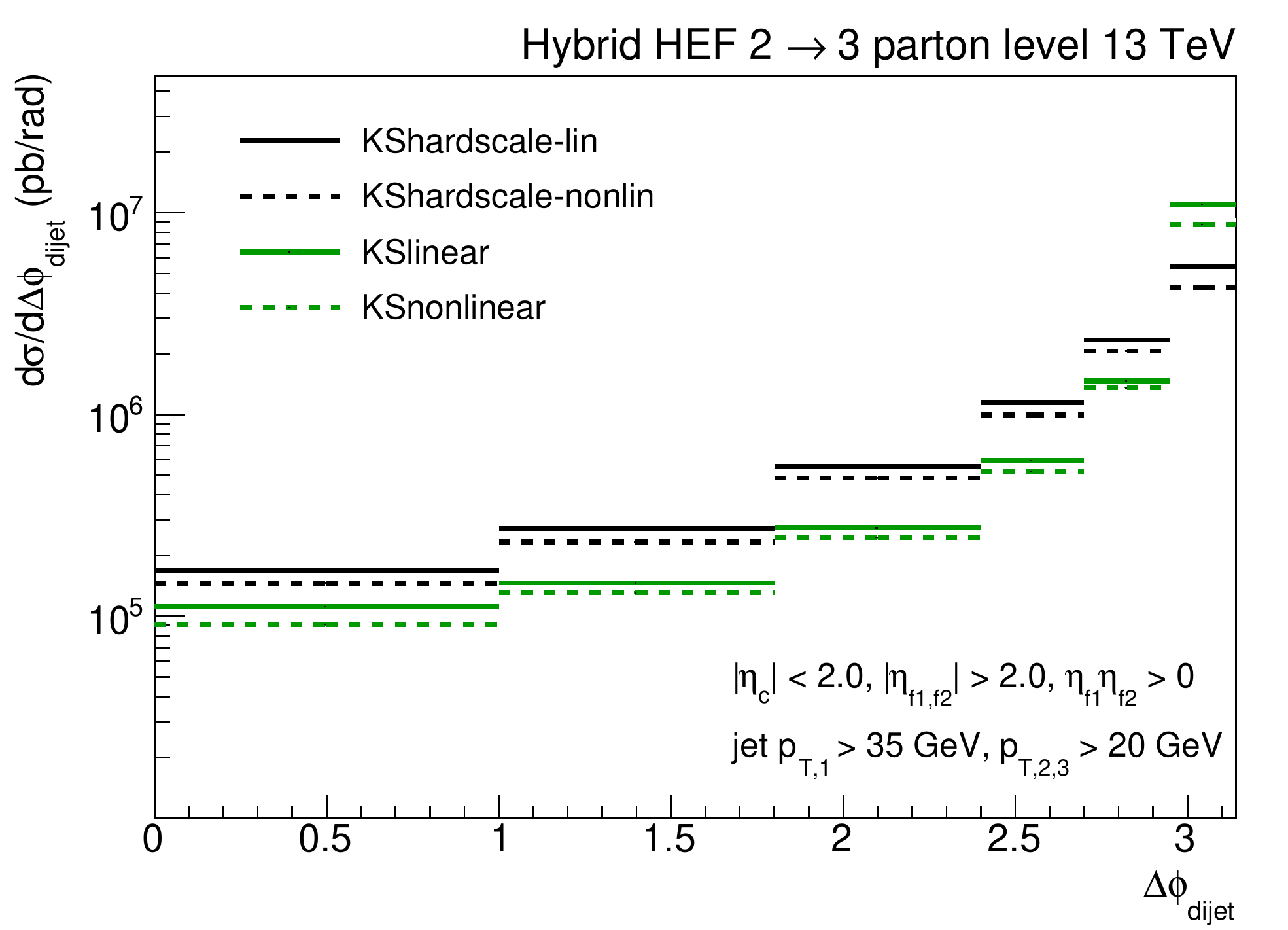}}
\subfloat[]{\includegraphics[width=0.52\textwidth]{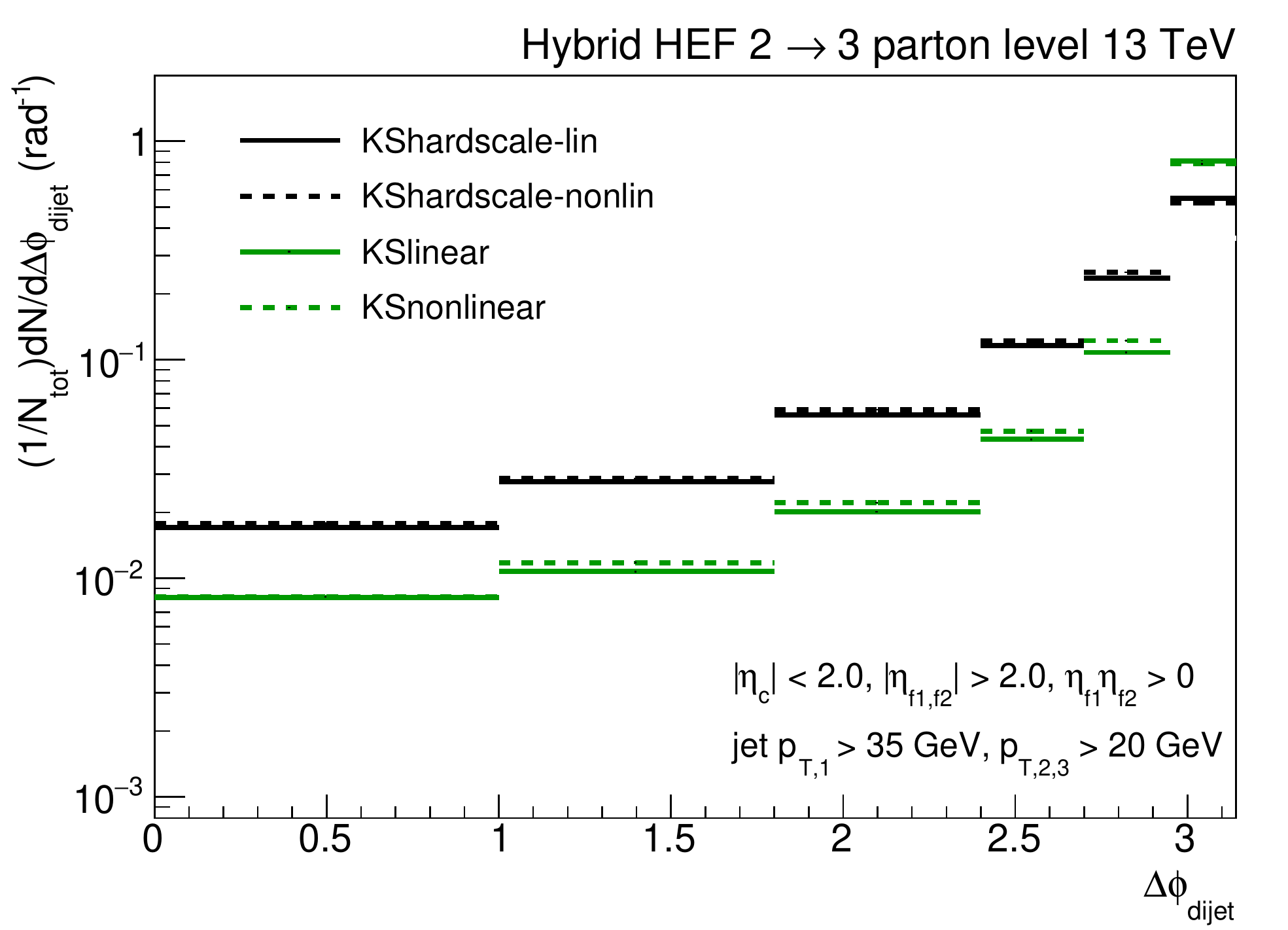}}}
\caption {Comparison of $2 \rightarrow 3$ parton level predictions using the hybrid framework with KS PDFs with (black lines) and without (green lines) Sudakov resummation. Shown in absolute cross sections (a) and normalised distributions (b).} 
\label{fig:KS_parton_level}
\end{figure}

We first compare the parton level predictions of the KS PDFs in the hybrid  framework. Figure \ref{fig:KS_parton_level} shows the azimuthal angle difference, $\Delta\phi_{\rm dijet}$, between the leading dijet system and the third jet, for both the linear and nonlinear PDFs with and without Sudakov resummation. The left figure (a) shows the absolute cross section predictions, while the right figure (b) illustrates the differences in shape by showing normalised distributions. These latter distributions are useful since it is known that the standard $\KT{}$-factorisation formula misses contributions from multiple partonic interactions, which mainly affect the normalisation. A dedicated study of these corrections has been done in \cite{Bury:2016cue,Kotko:2016lej}. In another recent study \cite{Blanco:2019qbm} it has been demonstrated that $\KT{}$-factorisation gives a good description of data when applied to purely colourless final states. The main difference observed is that the KS PDFs with Sudakov resummation result in a more flat shape of the spectrum with respect to the versions without it. The cross section becomes higher in the tail of the distribution towards $\Delta\phi_{\rm dijet}=0$, and is less peaked at $\Delta \phi_{\rm dijet}=\pi$. This happens because the Sudakov factor enhances contributions with larger incoming $\KT{}$, while the total cross section is roughly preserved. It thus suppresses strongest the configuration where the dijet system is balanced by the third jet, and it enhances the configuration where the angle between the considered final states is moderate.
In addition we see, especially in figure \ref{fig:KS_parton_level} (b), that there is no major difference between results based on linear and nonlinear PDFs for this observable and event topology. Therefore, this particular observable in the considered phase space is not sensitive to saturation effects and we can safely continue with the complete study. 

\begin{figure}[t!]                                   \centerline{\subfloat[]{\includegraphics[width=0.52\textwidth]{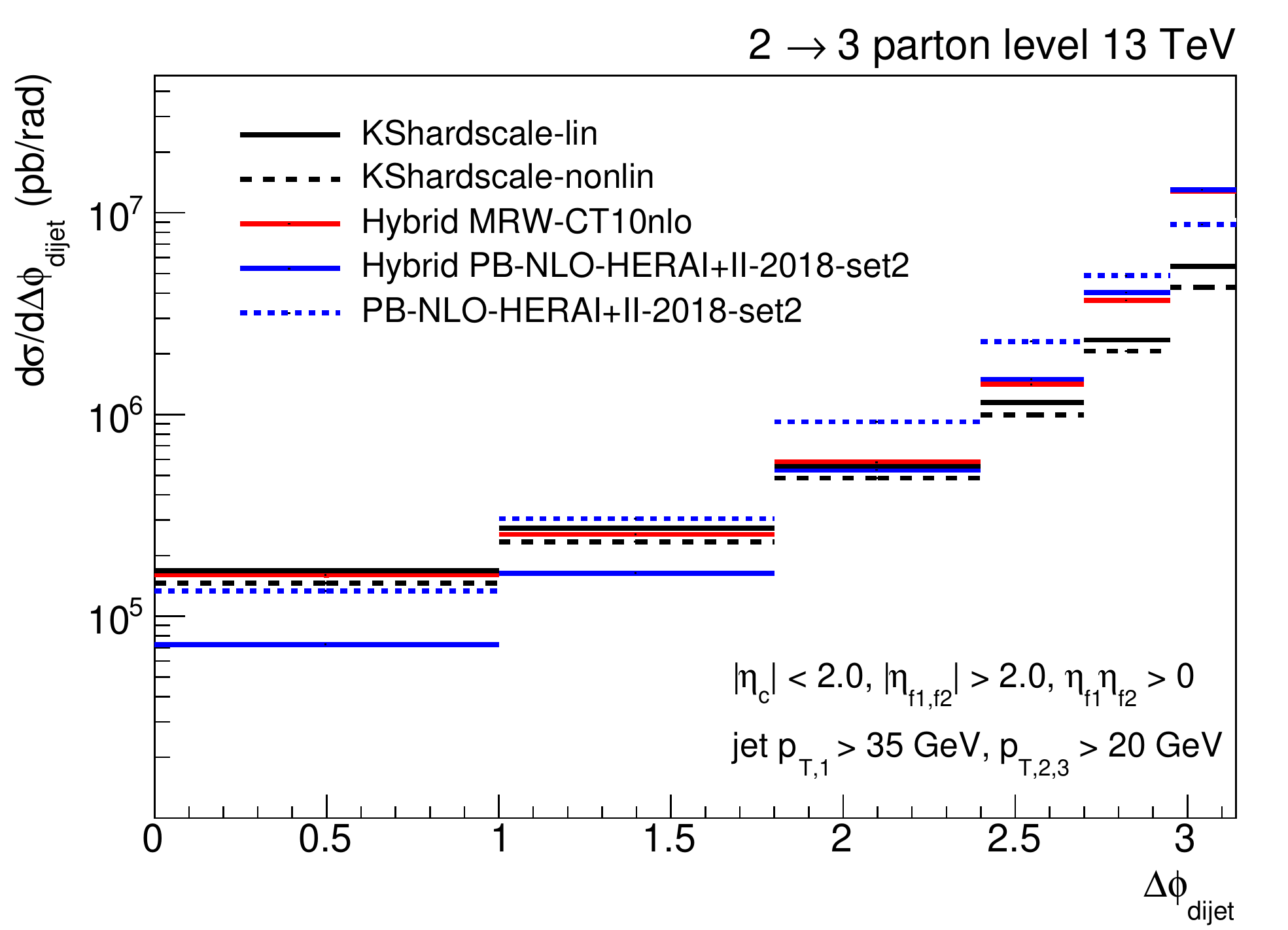}}
\subfloat[]{\includegraphics[width=0.52\textwidth]{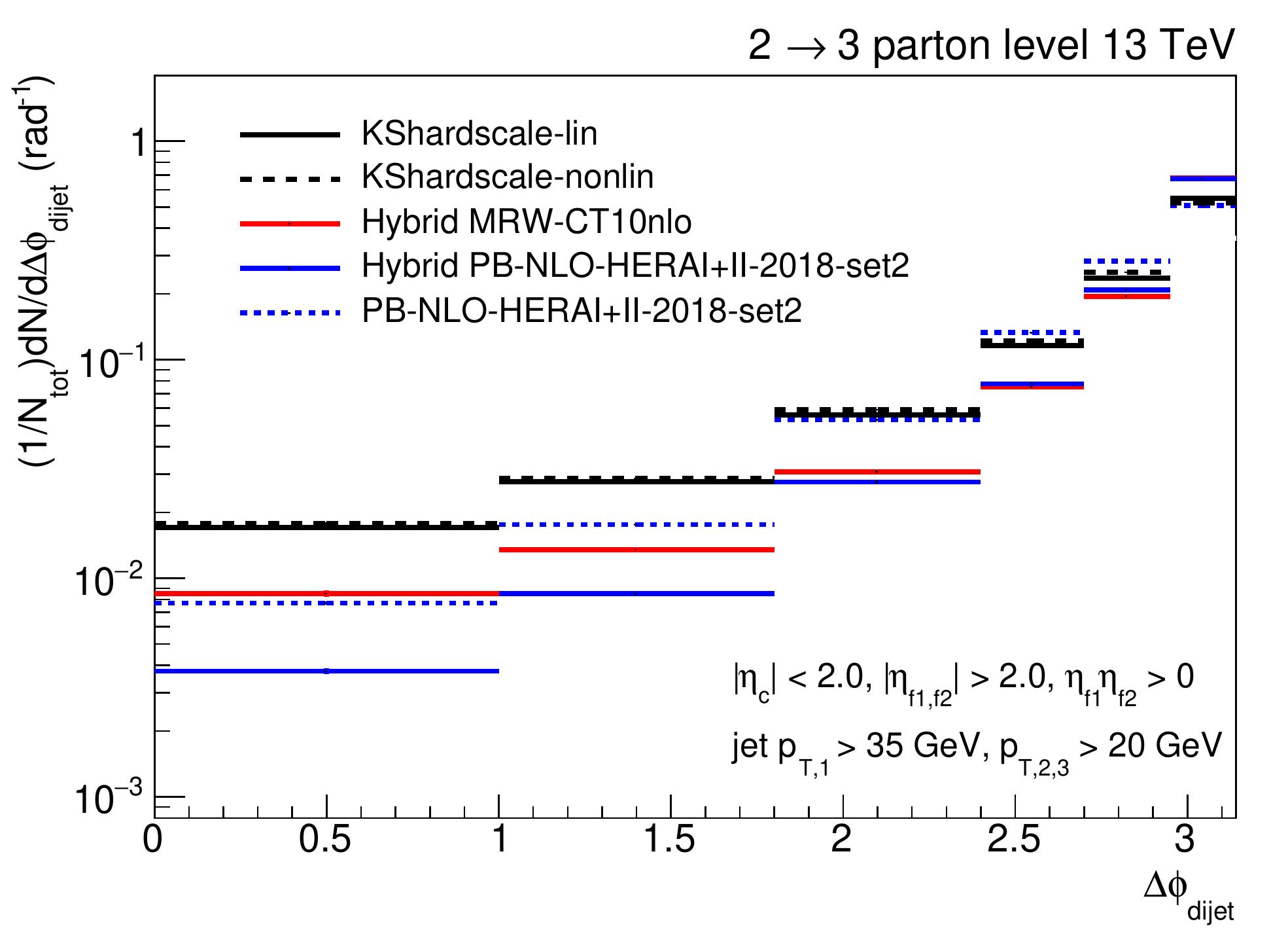}}}
\caption {Comparison of different $2 \rightarrow 3$ parton level predictions using both hybrid and full off-shell calculations. With KS PDFs including Sudakov resummation. Shown in absolute cross sections (a) and normalised distributions (b).} 
\label{fig:parton_level}
\end{figure}

\begin{figure}[t!]                                   \centerline{\subfloat[]{\includegraphics[width=0.52\textwidth]{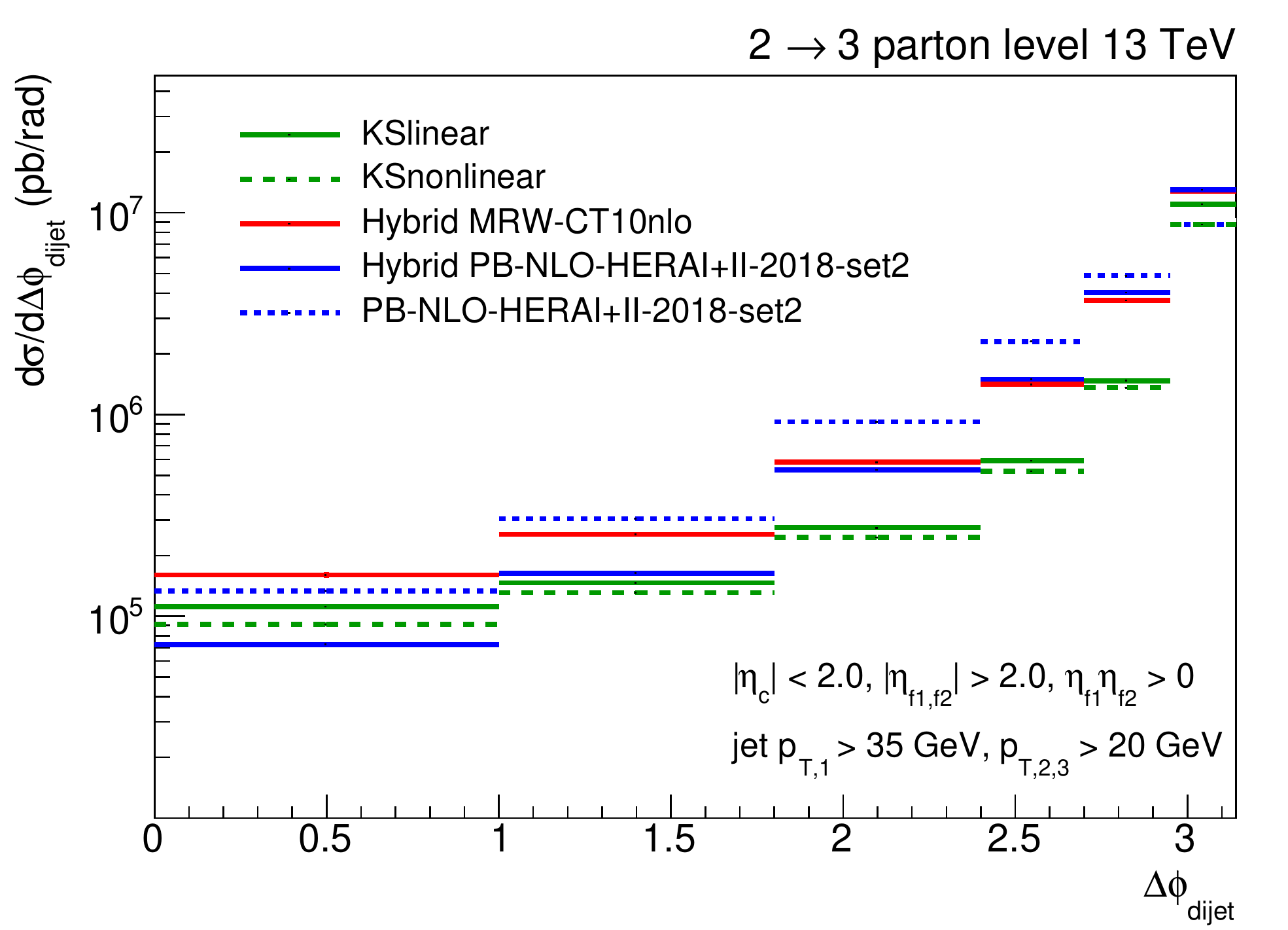}}
\subfloat[]{\includegraphics[width=0.52\textwidth]{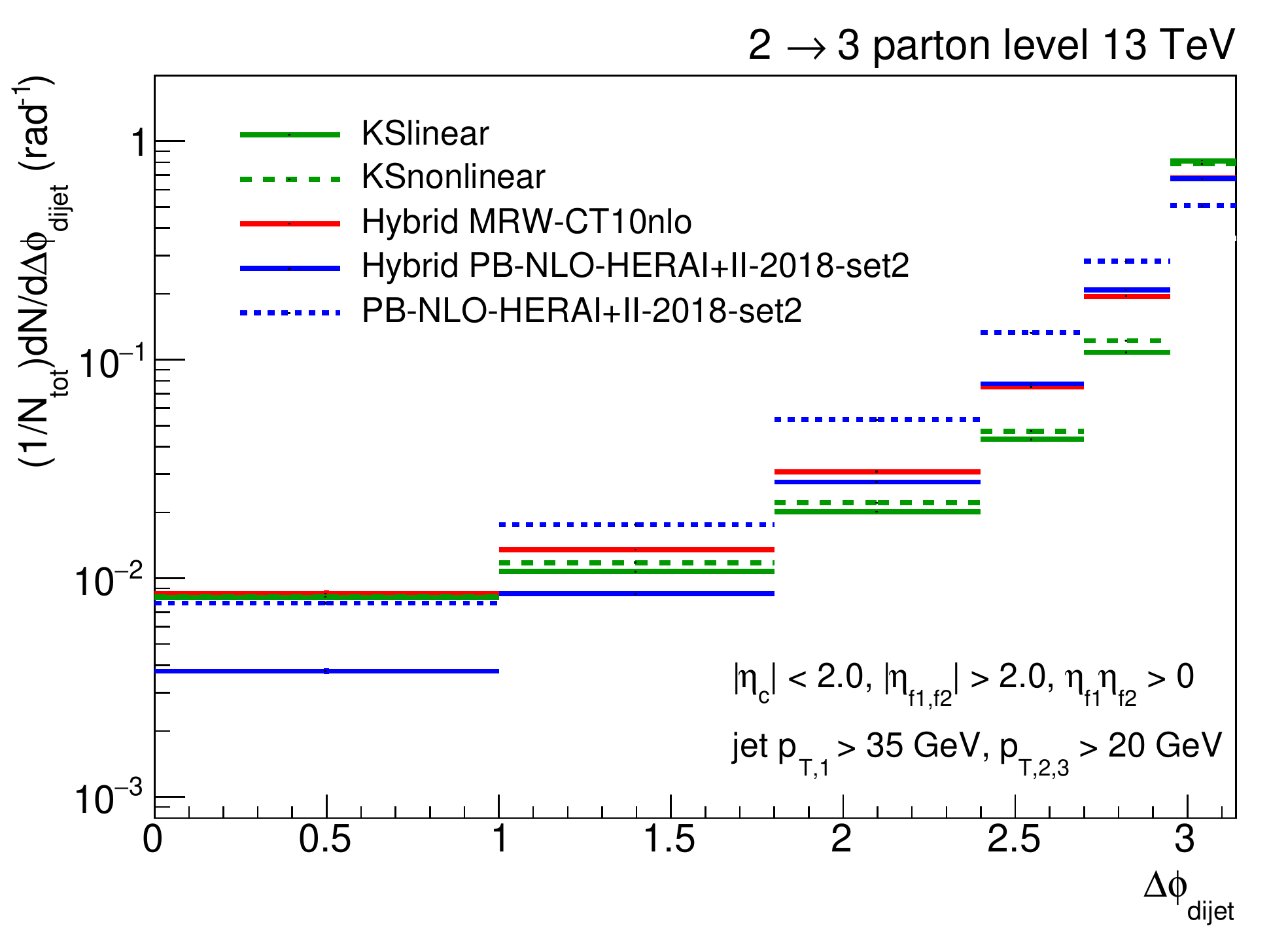}}}
\caption {Comparison of different $2 \rightarrow 3$ parton level predictions using both hybrid  and full off-shell calculations. With KS PDFs without Sudakov resummation. Shown in absolute cross sections (a) and normalised distributions (b).} 
\label{fig:parton_level_2}
\end{figure}

We can then extend the comparison by also including predictions obtained with the hybrid framework using the MRW-CT10nlo and PB-NLO-HERAI+II-2018-set2 full flavour PDFs. For the latter PDF we also include a prediction in which both initial partons have off-shell kinematics. 
Figures \ref{fig:parton_level} (a) and \ref{fig:parton_level_2} (a) show that the overall cross section is higher for the MRW-CT10nlo and PB-NLO-HERAI+II-2018-set2 PDFs compared to the KS PDFs used before. 
The reason for this is that KS PDFs were fitted with restriction to the low-$x$ data only while the other PDFs are valid in larger domain of $x$.
In addition, there is a difference between the hybrid and off-shell calculations using the same PB-NLO-HERAI+II-2018-set2 PDF: Figure \ref{fig:parton_level} (b) shows that the full off-shell curve is less peaked at $\Delta\phi_{\rm dijet} = \pi$. A nearly back-to-back configuration between the leading dijet system and third jet is less probable when two off-shell partons collide since the additional $\KT{}$ from the second unintegrated PDF increases the available phase space and allows for more decorrelation.
From this we conclude that the $\Delta\phi_{\rm dijet}$ observable has an excellent sensitivity to test both the applicability of the factorisation framework in a particular region of phase space, as well as to test and perhaps further constrain the PDFs used in the calculations.

\section{Effects of parton showers and hadronisation}
\label{sec:HadronLevel}

\begin{figure}[t!]                                   \centerline{\subfloat[]{\includegraphics[width=0.52\textwidth]{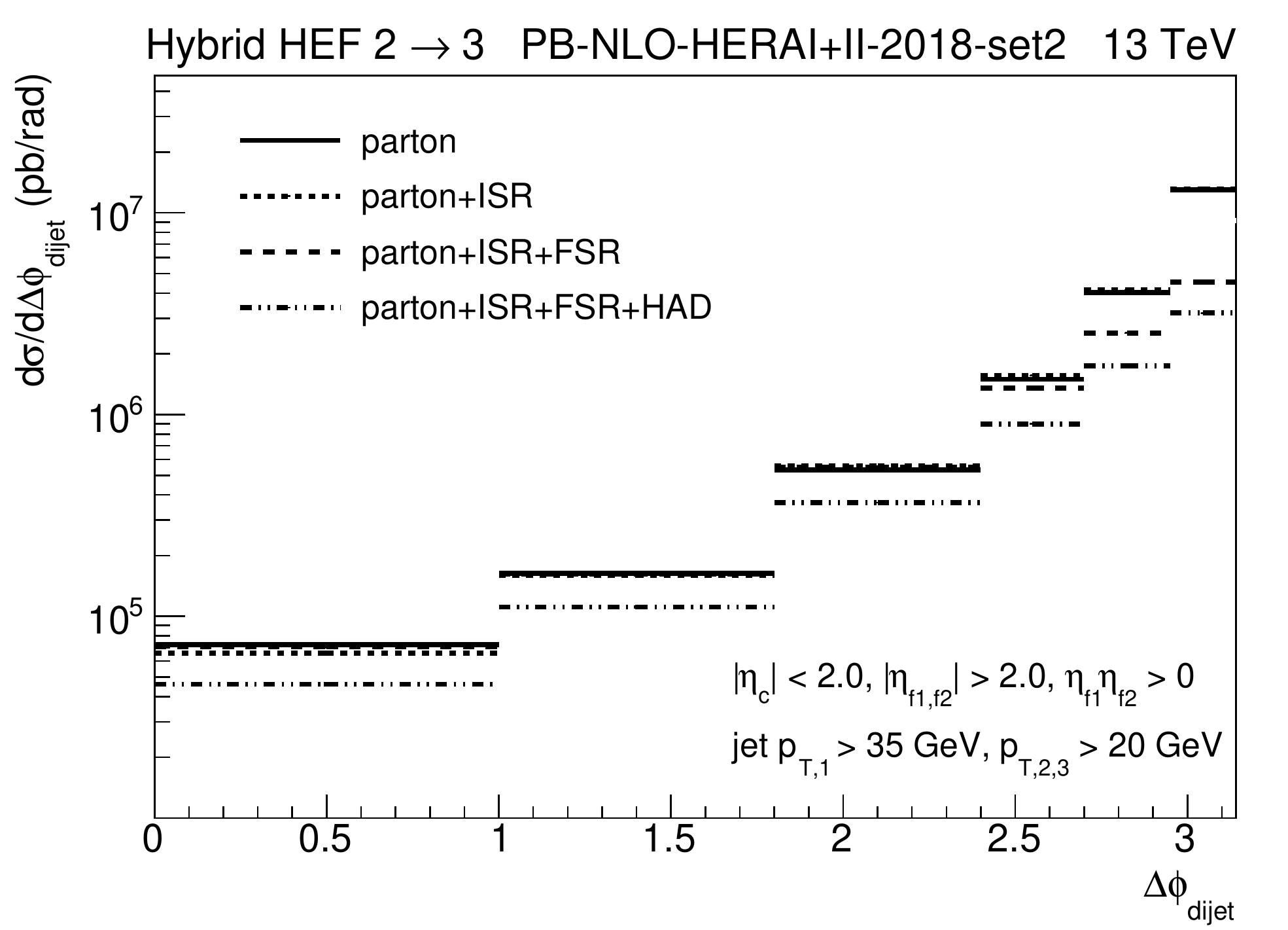}}
\subfloat[]{\includegraphics[width=0.52\textwidth]{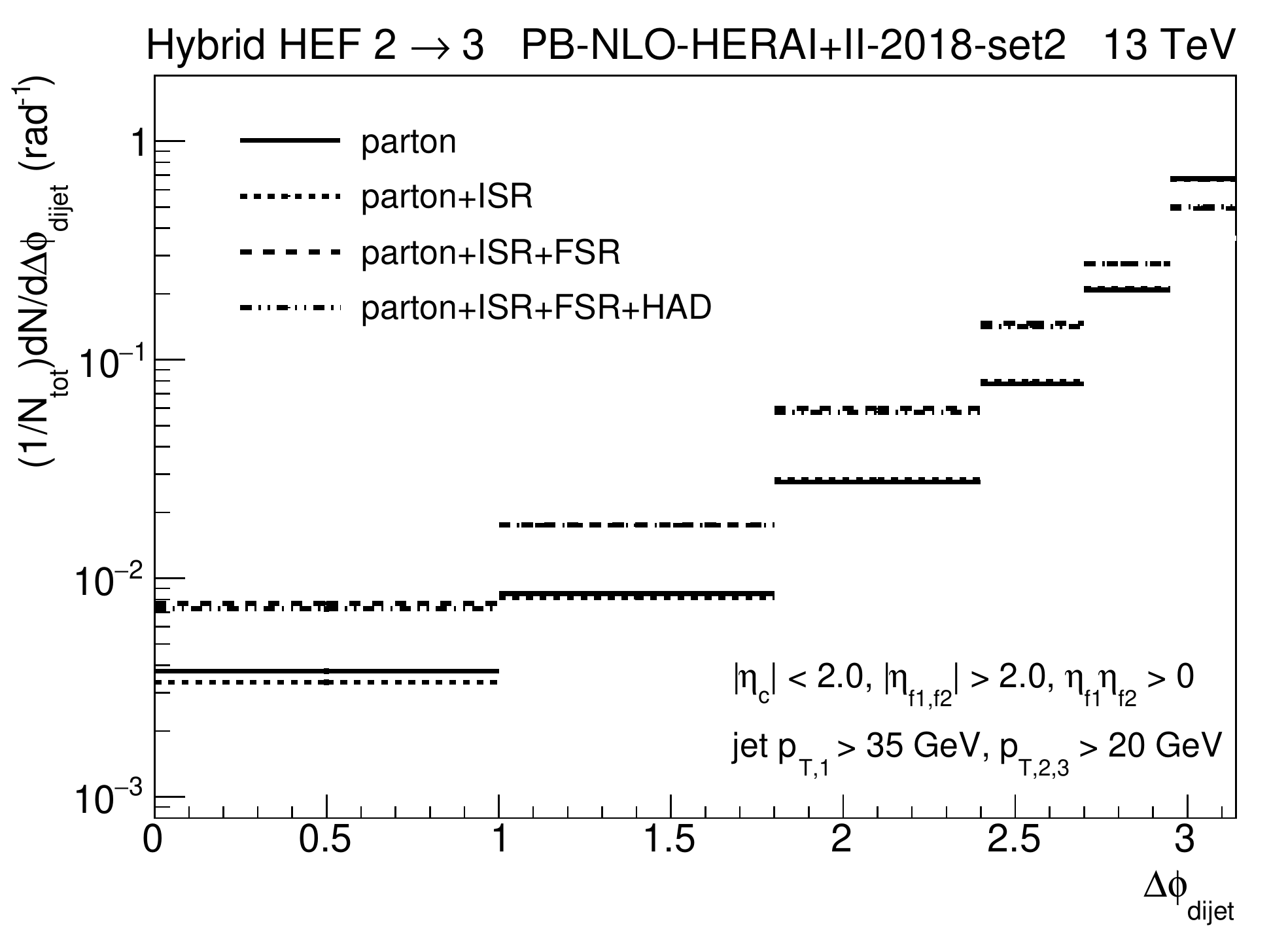}}}
\caption {Hybrid formalism predictions at parton level for $2 \rightarrow 3$ processes with subsequently adding ISR, FSR, and hadronisation. The PB-NLO-HERAI+II-2018-set2 PDF is used for all predictions. Shown in absolute cross sections (a) and normalised distributions (b).} 
\label{fig:parton_shower_effects}
\end{figure}

\begin{figure}[t!]                                   \centerline{\subfloat[]{\includegraphics[width=0.52\textwidth]{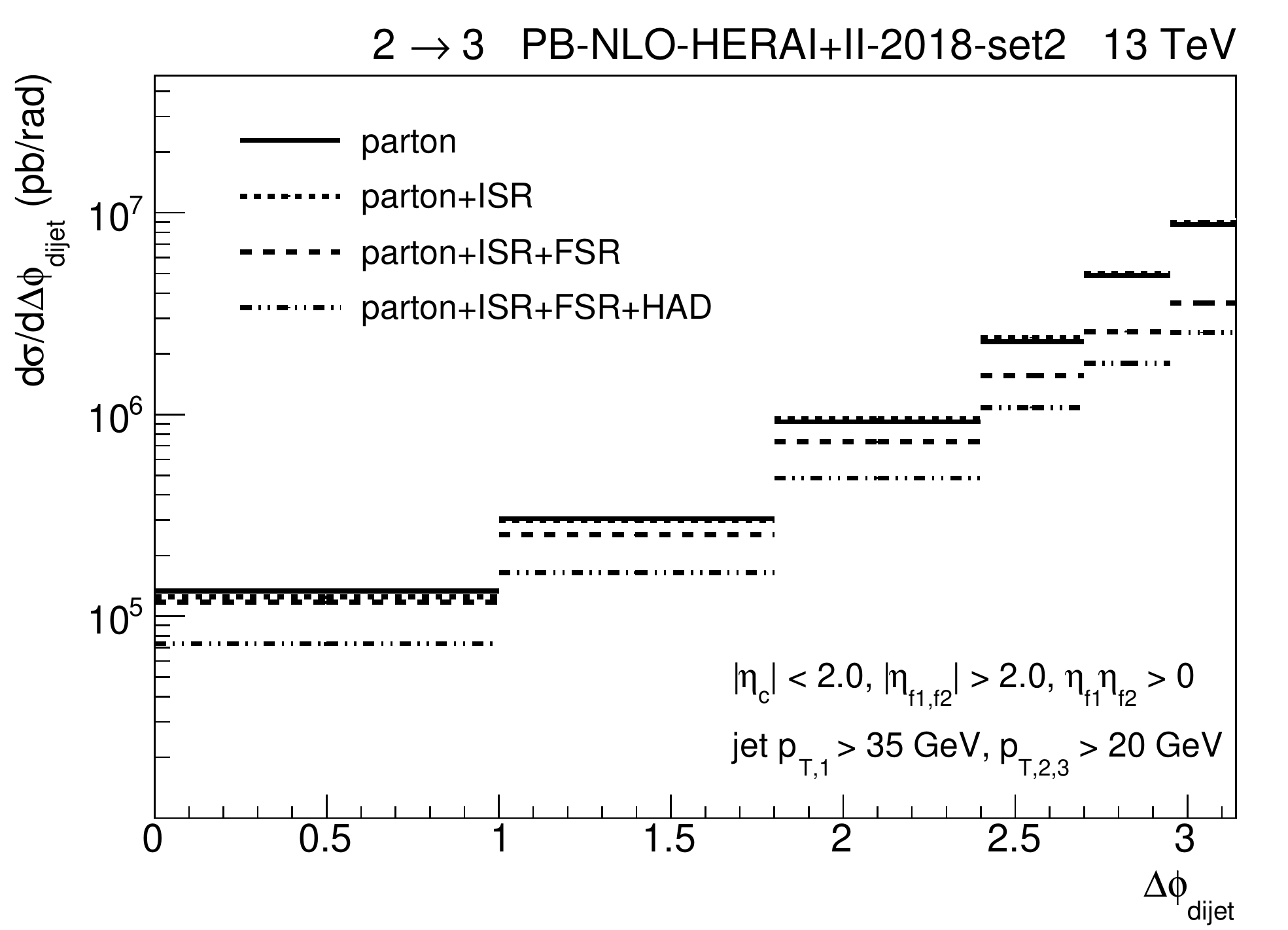}}
\subfloat[]{\includegraphics[width=0.52\textwidth]{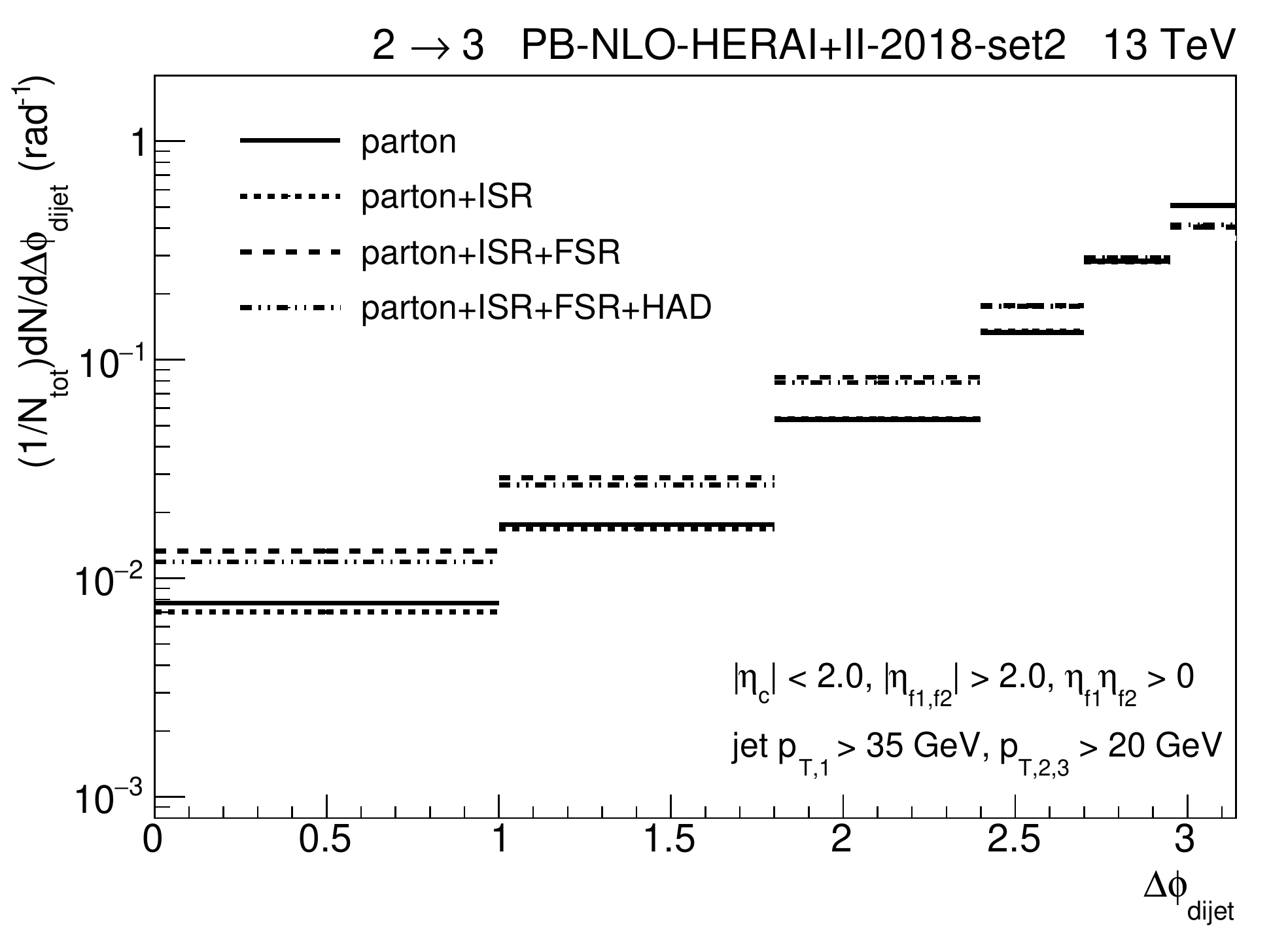}}}
\caption {Full off-shell predictions at parton level for $2 \rightarrow 3$ processes with subsequently adding ISR, FSR, and hadronisation. The PB-NLO-HERAI+II-2018-set2 PDF is used for all predictions. Shown in absolute cross sections (a) and normalised distributions (b).} 
\label{fig:parton_shower_effects_2}
\end{figure}

In this section we will investigate how the behaviour of the $\Delta\phi_{\rm dijet}$ observable changes when parton showers and hadronisation are added to the $2 \rightarrow 3$ process event generation. We do this for both the hybrid configuration and the full off-shell initial kinematics. To enable a consistent application of the PDF with parton shower effects, we use the PB-NLO-HERAI+II-2018-set2 parton branching unintegrated PDF. Figure \ref{fig:parton_shower_effects} shows the results for the hybrid formalism calculations (called later hybrid framework), and figure \ref{fig:parton_shower_effects_2} for the full off-shell configuration. The starting curve (solid line) shown in the figure represents the parton level results, and subsequently initial state radiation (short dashed line), final state radiation (long dashed line), and hadronisation (dash-dotted line) are added on top. 

The first observation that one can make is that there is basically no difference when adding ISR. This shows that the unintegrated parton density is consistent with initial state radiation, and one does not have to adjust kinematics in order to describe the final state. The convolution of a $2\rightarrow 3$ matrix element with an unintegrated parton density thus accounts for the bulk of kinematic effects. The second observation is that the situation changes when including final state radiation: the $\Delta\phi_{\rm dijet}$ distribution becomes less peaked, indicating an increased imbalance in the trijet system. This could be due to the radiation of partons outside the jet cone. Hadronisation, finally, results in an overall constant decrease of the cross section because the jet $\PT{}$ is lowered and falls below the imposed thresholds. In particular, figures \ref{fig:parton_shower_effects} (b) and \ref{fig:parton_shower_effects_2} (b) show the normalised predictions and confirm the aforementioned behaviour: only final state radiation causes a significant change in shape. These conclusions are both valid for the hybrid and full off-shell configurations.

\section{Effects of matrix element parton multiplicity}
\label{sec:Multiplicity}

In the previous sections, hard matrix elements for $2 \rightarrow 3$ processes were considered. In this section, we will additionally investigate $2 \rightarrow 2$ processes contributing to trijet final states, with one jet expected to come from the parton shower. The goal of this study is to determine in which region of the phase space one can approximate the full matrix element using a parton shower.  

\begin{figure}[t!]                                   \centerline{\subfloat[]{\includegraphics[width=0.49\textwidth]{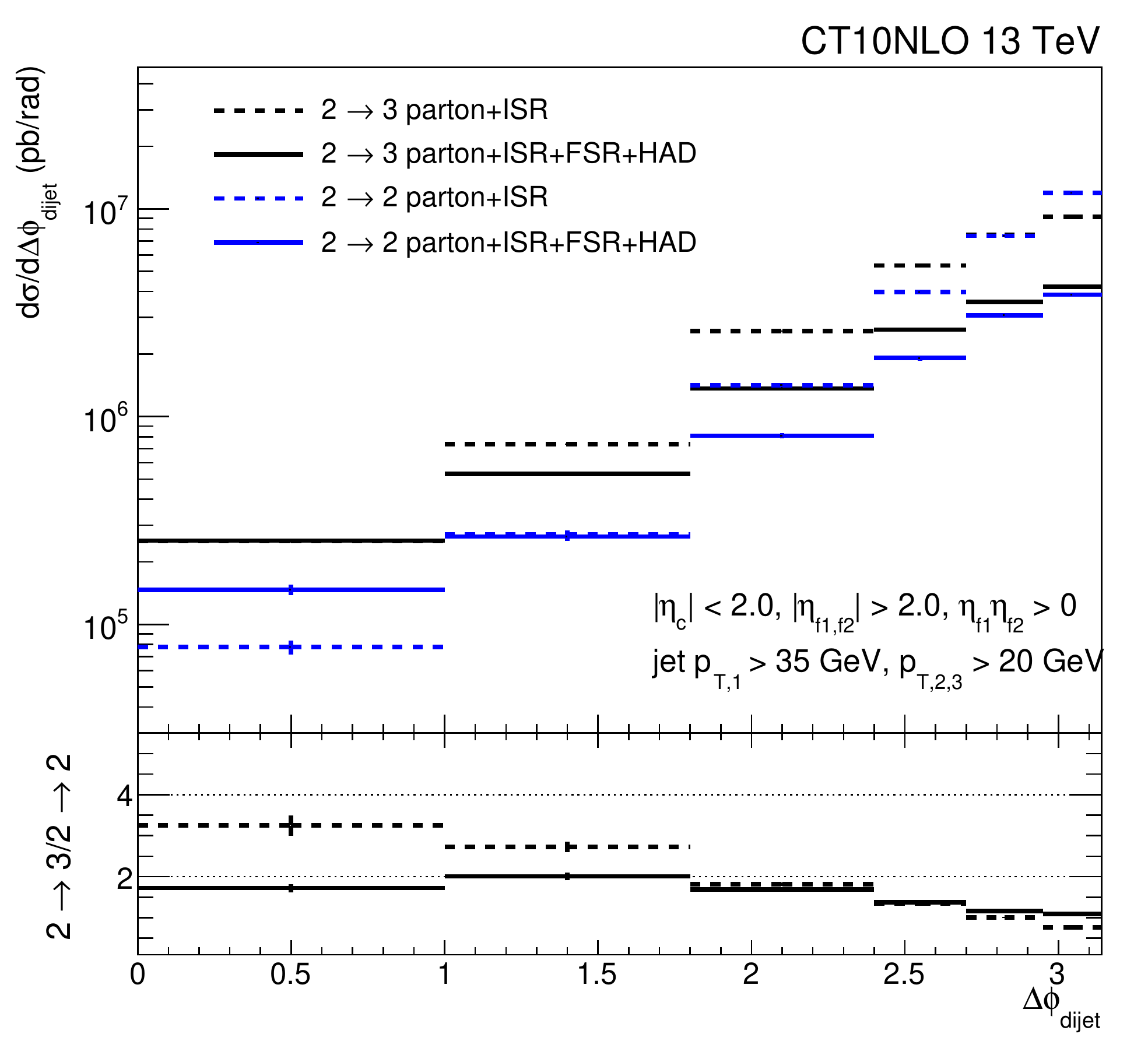}}
\subfloat[]{\includegraphics[width=0.49\textwidth]{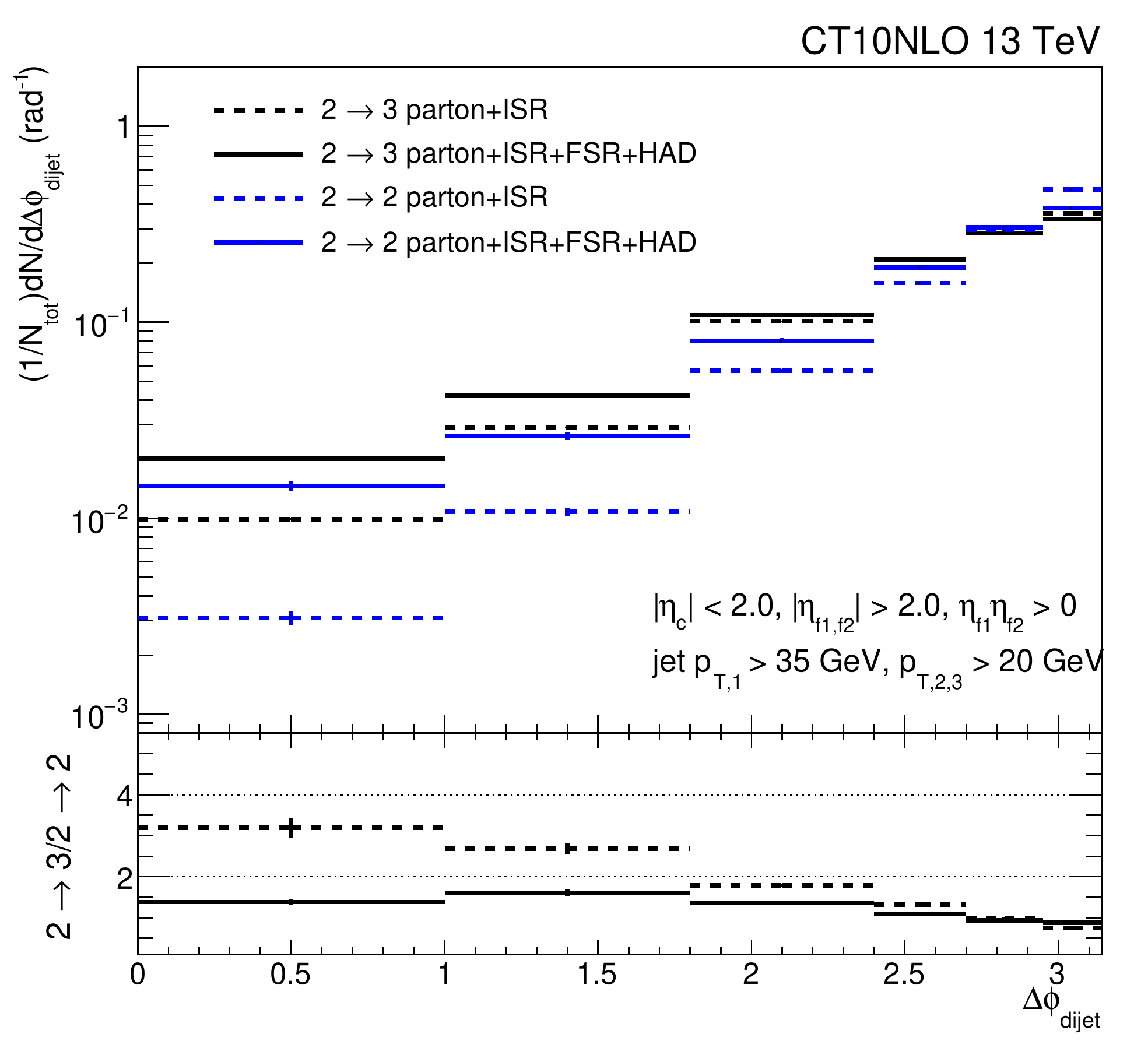}}}
\caption {On-shell predictions with the CT10NLO PDF with initial state radiation included (dashed lines), and at hadron level (solid lines) for both $2 \rightarrow 2$ (blue lines) and $2 \rightarrow 3$ (black lines) matrix element calculations. Shown in absolute cross sections (a) and normalised distributions (b). The bottom panel shows the ratio of the $2 \rightarrow 3$ over $2 \rightarrow 2$ predictions to illustrate the change in cross section.} 
\label{fig:ME_effects}
\end{figure}

Figure \ref{fig:ME_effects} shows the different configurations for on-shell calculations with the collinear CT10NLO PDF. The dashed lines present the results when only initial state radiation is included, while the solid lines show the results when also final state radiation and hadronisation are included. The black (blue) lines show the $2 \rightarrow 3$ ($2 \rightarrow 2$) processes.  One can see that when there are only 2 partons in the final state the $\Delta\phi_{\rm dijet}$ distribution is more peaked, indicating a smaller imbalance when one jet needs to come from the parton shower. 

\begin{figure}[t!]                                   \centerline{\subfloat[]{\includegraphics[width=0.49\textwidth]{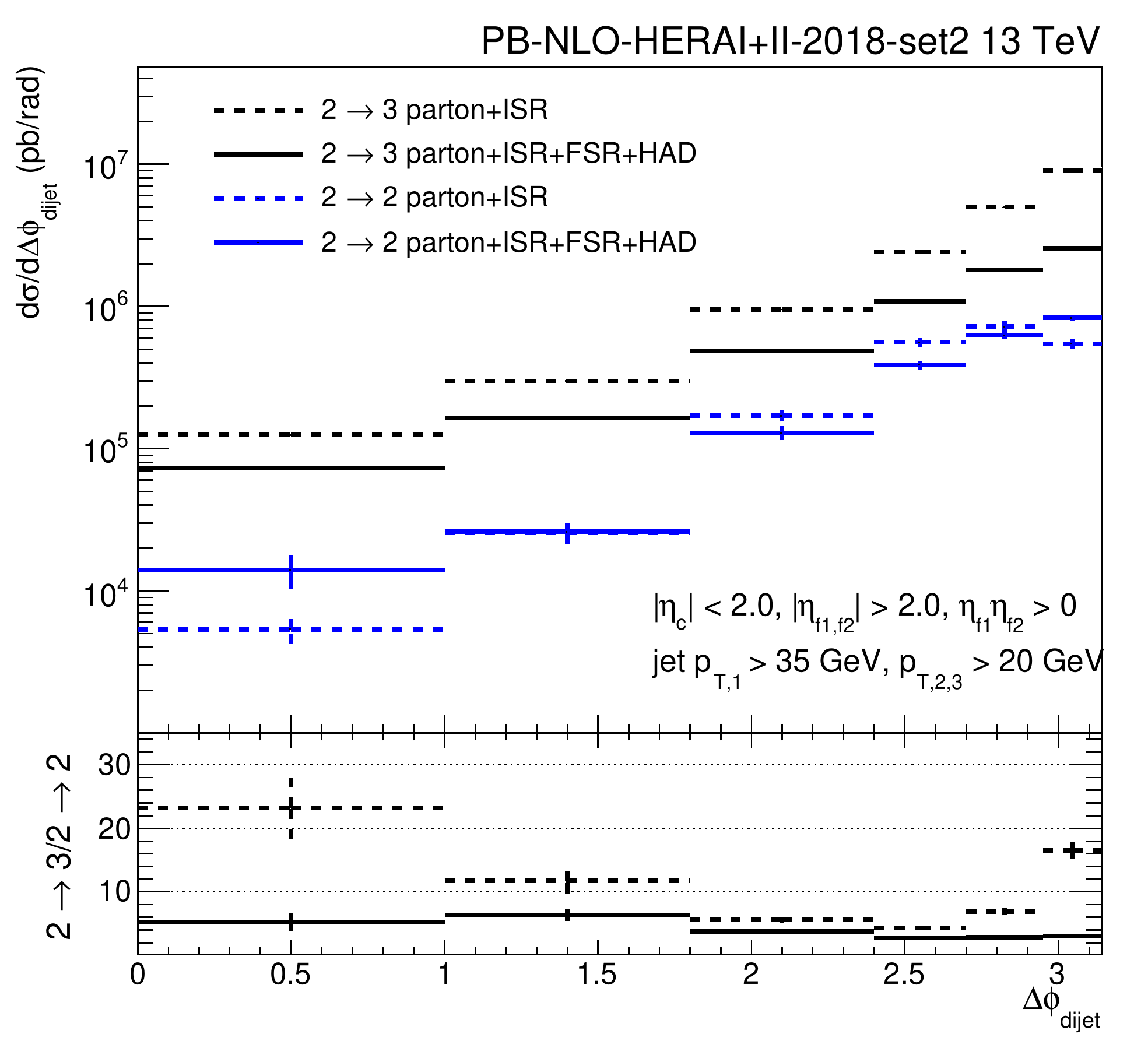}}
\subfloat[]{\includegraphics[width=0.49\textwidth]{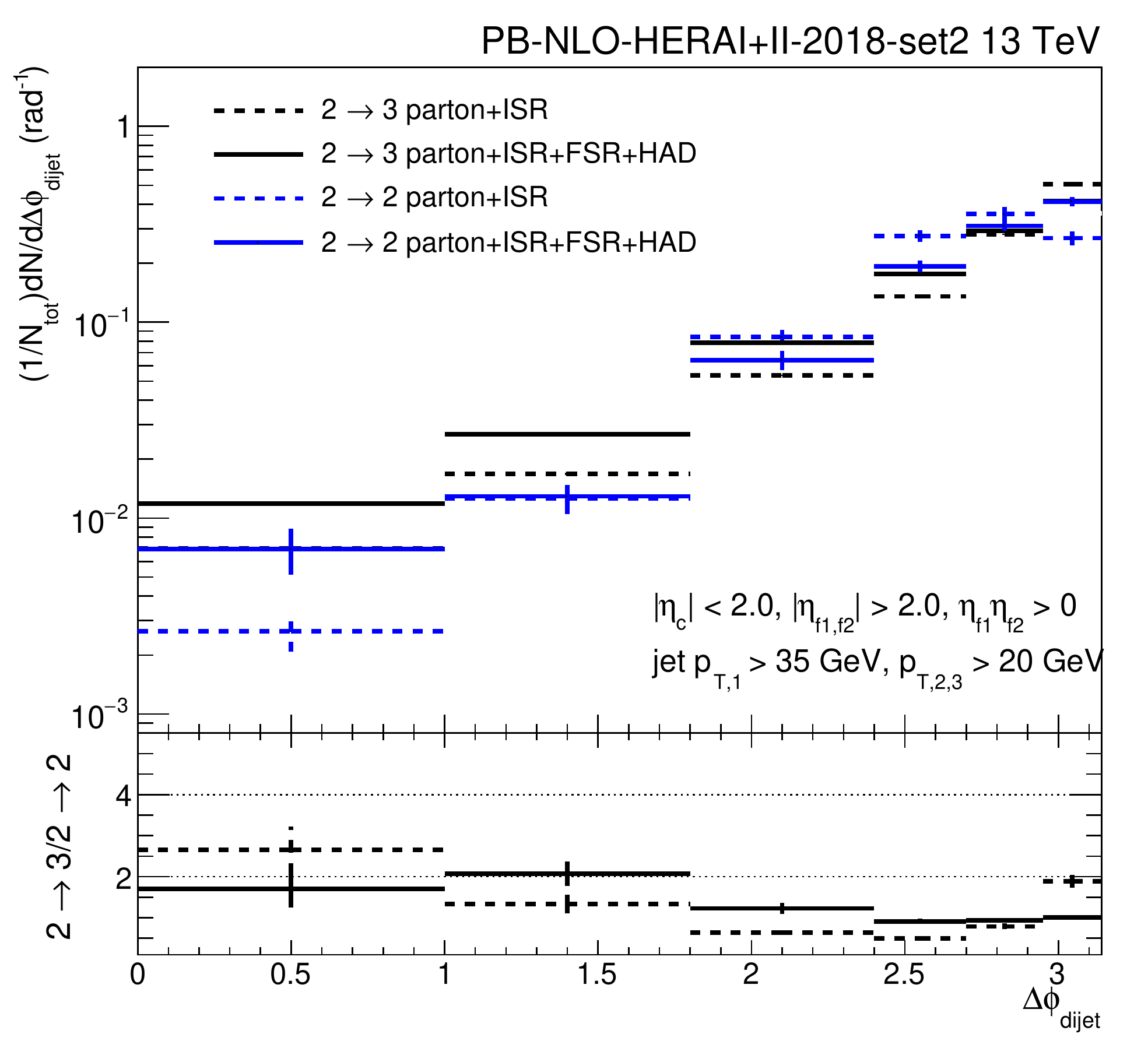}}}
\caption {Full off-shell predictions with the PB-NLO-HERAI+II-2018-set2 PDF, with initial state radiation included (dashed lines), and at hadron level (solid lines) for both $2 \rightarrow 2$ (blue lines) and $2 \rightarrow 3$ (black lines) matrix element calculations. Shown in absolute cross sections (a) and normalised distributions (b). The bottom panel shows the ratio of the $2 \rightarrow 3$ over $2 \rightarrow 2$ predictions to illustrate the change in cross section.} 
\label{fig:ME_effects_2}
\end{figure}

Figure \ref{fig:ME_effects_2} shows the same content but with off-shell calculations using the parton branching unintegrated PDF. In this case a larger difference in cross section between the $2 \rightarrow 3$ and $2 \rightarrow 2$ processes is visible (shown clearly in the ratio panels of figures \ref{fig:ME_effects} (a) and \ref{fig:ME_effects_2} (a)). The cross section of the latter configuration is significantly lower, and the effect of adding final state radiation and hadronisation leads to a similar result with only a small difference towards $\Delta\phi_{\rm dijet} = \pi$. This is in contrast to the $2 \rightarrow 3$ processes where adding FSR and hadronisation effects clearly lower the cross section. The lower cross section of the $2 \rightarrow 2$ processes could imply that the $\PT{}$ of the jets generated in the initial state parton shower is on average too low to pass the analysis cuts. Depending on how these curves would describe a measurement with data, it might thus be needed to further fine tune ISR within the parton branching method.

As a result, the $\Delta\phi_{\rm dijet}$ of this particular 3-jet event topology is ideal to study the performance of different types of parton showers, and a measurement can help to constrain the expected jet cross sections.

\section{Conclusions}
In this paper we studied 3-jet production in proton-proton collisions at an LHC energy of $\sqrt{s}$ = 13 TeV. As a theoretical tool we used $\KT{}$-factorisation and the hybrid framework implemented in Monte Carlo event generators: \katie\ for the matrix element generation and \cascade\ for the parton shower development. The proposed final state, i.e.\ three jets in a central-forward configuration, and the $\Delta\phi_{\rm dijet}$ observable that describes the azimuthal angle difference between the leading dijet system and the third jet are ideal to study the performance of different collinear and unintegrated PDFs with on-shell, hybrid formalism, or off-shell calculations. It is furthermore well suited to study the effects of parton showers.

It is confirmed that the topology is not sensitive to nonlinear gluon density effects, and it is shown that there is a large difference between predictions of $2 \rightarrow 3$ and $2 \rightarrow 2$ processes at hadron level when two initial off-shell partons are used in the calculations. Finally, it is also confirmed that the discriminating power of the $\Delta\phi_{\rm dijet}$ observable remains after including parton showers and hadronisation, i.e.\ after taking non-perturbative corrections into account.

A measurement of the discussed 3-jet event topology in proton-proton collisions data at a centre-of-mass energy of $\sqrt{s}$ = 13 TeV can thus yield important information to improve the current available theoretical frameworks.

\label{sec:Summary}

\section{Acknowledgements}
 This work is supported partially by grant V03319N from the common FWO-PAS exchange program. Krzysztof Kutak acknowledges the DAAD fellowship "Research Stays for University Academics and Scientists", 2019 (57440915) to stay at DESY, during which part of this work was done and Hannes Jung for many valuable discussions.
 PK is supported by the National Science Centre, Poland, grant  2018/31/D/ST2/02731.

\bibliographystyle{JHEP}
\bibliography{references}

\providecommand{\href}[2]{#2}\begingroup\raggedright\begin{thebibliography}{10}

\bibitem{Sapeta:2015gee}
S.~Sapeta, \emph{{QCD and Jets at Hadron Colliders}},
  \href{http://dx.doi.org/10.1016/j.ppnp.2016.02.002}{\emph{Prog. Part. Nucl.
  Phys.} {\bf 89} (2016) 1--55}, [\href{https://arxiv.org/abs/1511.09336}{{\tt
  1511.09336}}].

\bibitem{Collins:2011zzd}
J.~Collins, \emph{{Foundations of perturbative QCD}}, vol.~32.
\newblock Cambridge Univ. Press, 2011.

\bibitem{Collins:1984kg}
J.~C. Collins, D.~E. Soper and G.~F. Sterman, \emph{{Transverse Momentum
  Distribution in Drell-Yan Pair and W and Z Boson Production}},
  \href{http://dx.doi.org/10.1016/0550-3213(85)90479-1}{\emph{Nucl. Phys.} {\bf
  B250} (1985) 199--224}.

\bibitem{Scimemi:2019cmh}
I.~Scimemi and A.~Vladimirov, \emph{{Non-perturbative structure of
  semi-inclusive deep-inelastic and Drell-Yan scattering at small transverse
  momentum}},  \href{https://arxiv.org/abs/1912.06532}{{\tt 1912.06532}}.

\bibitem{Bertone:2019nxa}
V.~Bertone, I.~Scimemi and A.~Vladimirov, \emph{{Extraction of unpolarized
  quark transverse momentum dependent parton distributions from
  Drell-Yan/Z-boson production}},
  \href{http://dx.doi.org/10.1007/JHEP06(2019)028}{\emph{JHEP} {\bf 06} (2019)
  028}, [\href{https://arxiv.org/abs/1902.08474}{{\tt 1902.08474}}].

\bibitem{Bacchetta:2019sam}
A.~Bacchetta, V.~Bertone, C.~Bissolotti, G.~Bozzi, F.~Delcarro, F.~Piacenza
  et~al., \emph{{Transverse-momentum-dependent parton distributions up to
  N$^3$LL from Drell-Yan data}},  \href{https://arxiv.org/abs/1912.07550}{{\tt
  1912.07550}}.

\bibitem{Collins:1991ty}
J.~C. Collins and R.~K. Ellis, \emph{{Heavy quark production in very
  high-energy hadron collisions}},
  \href{http://dx.doi.org/10.1016/0550-3213(91)90288-9}{\emph{Nucl. Phys.} {\bf
  B360} (1991) 3--30}.

\bibitem{Catani:1990eg}
S.~Catani, M.~Ciafaloni and F.~Hautmann, \emph{{High-energy factorization and
  small x heavy flavor production}},
  \href{http://dx.doi.org/10.1016/0550-3213(91)90055-3}{\emph{Nucl. Phys.} {\bf
  B366} (1991) 135--188}.

\bibitem{Sjostrand:2006za}
T.~Sjostrand, S.~Mrenna and P.~Z. Skands, \emph{{PYTHIA 6.4 Physics and
  Manual}}, \href{http://dx.doi.org/10.1088/1126-6708/2006/05/026}{\emph{JHEP}
  {\bf 05} (2006) 026}, [\href{https://arxiv.org/abs/hep-ph/0603175}{{\tt
  hep-ph/0603175}}].

\bibitem{Sjostrand:2014zea}
T.~Sjöstrand, S.~Ask, J.~R. Christiansen, R.~Corke, N.~Desai, P.~Ilten et~al.,
  \emph{{An Introduction to PYTHIA 8.2}},
  \href{http://dx.doi.org/10.1016/j.cpc.2015.01.024}{\emph{Comput. Phys.
  Commun.} {\bf 191} (2015) 159--177},
  [\href{https://arxiv.org/abs/1410.3012}{{\tt 1410.3012}}].

\bibitem{Bahr:2008pv}
M.~Bahr et~al., \emph{{Herwig++ Physics and Manual}},
  \href{http://dx.doi.org/10.1140/epjc/s10052-008-0798-9}{\emph{Eur. Phys. J.}
  {\bf C58} (2008) 639--707}, [\href{https://arxiv.org/abs/0803.0883}{{\tt
  0803.0883}}].

\bibitem{Bellm:2015jjp}
J.~Bellm et~al., \emph{{Herwig 7.0/Herwig++ 3.0 release note}},
  \href{http://dx.doi.org/10.1140/epjc/s10052-016-4018-8}{\emph{Eur.\ Phys.\
  J.\ C} {\bf 76} (2016) 196}, [\href{https://arxiv.org/abs/1512.01178}{{\tt
  1512.01178}}].

\bibitem{Bothmann:2019yzt}
{\scshape Sherpa} collaboration, E.~Bothmann et~al., \emph{{Event Generation
  with Sherpa 2.2}},
  \href{http://dx.doi.org/10.21468/SciPostPhys.7.3.034}{\emph{SciPost Phys.}
  {\bf 7} (2019) 034}, [\href{https://arxiv.org/abs/1905.09127}{{\tt
  1905.09127}}].

\bibitem{Bury:2017jxo}
M.~Bury, A.~van Hameren, H.~Jung, K.~Kutak, S.~Sapeta and M.~Serino,
  \emph{{Calculations with off-shell matrix elements, TMD parton densities and
  TMD parton showers}},
  \href{http://dx.doi.org/10.1140/epjc/s10052-018-5642-2}{\emph{Eur. Phys. J.}
  {\bf C78} (2018) 137}, [\href{https://arxiv.org/abs/1712.05932}{{\tt
  1712.05932}}].

\bibitem{Fadin:1975cb}
V.~S. Fadin, E.~A. Kuraev and L.~N. Lipatov, \emph{{On the Pomeranchuk
  Singularity in Asymptotically Free Theories}},
  \href{http://dx.doi.org/10.1016/0370-2693(75)90524-9}{\emph{Phys. Lett.} {\bf
  60B} (1975) 50--52}.

\bibitem{Balitsky:1978ic}
I.~I. Balitsky and L.~N. Lipatov, \emph{{The Pomeranchuk Singularity in Quantum
  Chromodynamics}}, {\emph{Sov. J. Nucl. Phys.} {\bf 28} (1978) 822--829}.

\bibitem{Kimber:2001sc}
M.~A. Kimber, A.~D. Martin and M.~G. Ryskin, \emph{{Unintegrated parton
  distributions}},
  \href{http://dx.doi.org/10.1103/PhysRevD.63.114027}{\emph{Phys. Rev.} {\bf
  D63} (2001) 114027}, [\href{https://arxiv.org/abs/hep-ph/0101348}{{\tt
  hep-ph/0101348}}].

\bibitem{vanHameren:2014ala}
A.~van Hameren, P.~Kotko, K.~Kutak and S.~Sapeta, \emph{{Small-$x$ dynamics in
  forward-central dijet decorrelations at the LHC}},
  \href{http://dx.doi.org/10.1016/j.physletb.2014.09.005}{\emph{Phys. Lett.}
  {\bf B737} (2014) 335--340}, [\href{https://arxiv.org/abs/1404.6204}{{\tt
  1404.6204}}].

\bibitem{Kutak:2012rf}
K.~Kutak and S.~Sapeta, \emph{{Gluon saturation in dijet production in p-Pb
  collisions at Large Hadron Collider}},
  \href{http://dx.doi.org/10.1103/PhysRevD.86.094043}{\emph{Phys. Rev.} {\bf
  D86} (2012) 094043}, [\href{https://arxiv.org/abs/1205.5035}{{\tt
  1205.5035}}].

\bibitem{Watt:2003mx}
G.~Watt, A.~D. Martin and M.~G. Ryskin, \emph{{Unintegrated parton
  distributions and inclusive jet production at HERA}},
  \href{http://dx.doi.org/10.1140/epjc/s2003-01320-4}{\emph{Eur. Phys. J.} {\bf
  C31} (2003) 73--89}, [\href{https://arxiv.org/abs/hep-ph/0306169}{{\tt
  hep-ph/0306169}}].

\bibitem{Antonov:2004hh}
E.~N. Antonov, L.~N. Lipatov, E.~A. Kuraev and I.~O. Cherednikov,
  \emph{{Feynman rules for effective Regge action}},
  \href{http://dx.doi.org/10.1016/j.nuclphysb.2005.05.013,
  10.1016/j.nuclphysb.2005.013}{\emph{Nucl. Phys.} {\bf B721} (2005) 111--135},
  [\href{https://arxiv.org/abs/hep-ph/0411185}{{\tt hep-ph/0411185}}].

\bibitem{vanHameren:2012if}
A.~van Hameren, P.~Kotko and K.~Kutak, \emph{{Helicity amplitudes for
  high-energy scattering}},
  \href{http://dx.doi.org/10.1007/JHEP01(2013)078}{\emph{JHEP} {\bf 01} (2013)
  078}, [\href{https://arxiv.org/abs/1211.0961}{{\tt 1211.0961}}].

\bibitem{vanHameren:2012uj}
A.~van Hameren, P.~Kotko and K.~Kutak, \emph{{Multi-gluon helicity amplitudes
  with one off-shell leg within high energy factorization}},
  \href{http://dx.doi.org/10.1007/JHEP12(2012)029}{\emph{JHEP} {\bf 12} (2012)
  029}, [\href{https://arxiv.org/abs/1207.3332}{{\tt 1207.3332}}].

\bibitem{Kotko:2014aba}
P.~Kotko, \emph{{Wilson lines and gauge invariant off-shell amplitudes}},
  \href{http://dx.doi.org/10.1007/JHEP07(2014)128}{\emph{JHEP} {\bf 07} (2014)
  128}, [\href{https://arxiv.org/abs/1403.4824}{{\tt 1403.4824}}].

\bibitem{vanHameren:2015bba}
A.~van Hameren and M.~Serino, \emph{{BCFW recursion for TMD parton
  scattering}}, \href{http://dx.doi.org/10.1007/JHEP07(2015)010}{\emph{JHEP}
  {\bf 07} (2015) 010}, [\href{https://arxiv.org/abs/1504.00315}{{\tt
  1504.00315}}].

\bibitem{Deak:2010gk}
M.~Deak, F.~Hautmann, H.~Jung and K.~Kutak, \emph{{Forward-Central Jet
  Correlations at the Large Hadron Collider}},
  \href{https://arxiv.org/abs/1012.6037}{{\tt 1012.6037}}.

\bibitem{Deak:2011ga}
M.~Deak, F.~Hautmann, H.~Jung and K.~Kutak, \emph{{Forward Jets and Energy Flow
  in Hadronic Collisions}},
  \href{http://dx.doi.org/10.1140/epjc/s10052-012-1982-5}{\emph{Eur. Phys. J.
  C} {\bf 72} (2012) 1982}, [\href{https://arxiv.org/abs/1112.6354}{{\tt
  1112.6354}}].

\bibitem{Gribov:1984tu}
L.~V. Gribov, E.~M. Levin and M.~G. Ryskin, \emph{{Semihard Processes in QCD}},
  \href{http://dx.doi.org/10.1016/0370-1573(83)90022-4}{\emph{Phys. Rept.} {\bf
  100} (1983) 1--150}.

\bibitem{McLerran:1993ka}
L.~D. McLerran and R.~Venugopalan, \emph{{Gluon distribution functions for very
  large nuclei at small transverse momentum}},
  \href{http://dx.doi.org/10.1103/PhysRevD.49.3352}{\emph{Phys. Rev.} {\bf D49}
  (1994) 3352--3355}, [\href{https://arxiv.org/abs/hep-ph/9311205}{{\tt
  hep-ph/9311205}}].

\bibitem{Kovchegov:1999yj}
Y.~V. Kovchegov, \emph{{Small x F(2) structure function of a nucleus including
  multiple pomeron exchanges}},
  \href{http://dx.doi.org/10.1103/PhysRevD.60.034008}{\emph{Phys. Rev.} {\bf
  D60} (1999) 034008}, [\href{https://arxiv.org/abs/hep-ph/9901281}{{\tt
  hep-ph/9901281}}].

\bibitem{Balitsky:1995ub}
I.~Balitsky, \emph{{Operator expansion for high-energy scattering}},
  \href{http://dx.doi.org/10.1016/0550-3213(95)00638-9}{\emph{Nucl. Phys.} {\bf
  B463} (1996) 99--160}, [\href{https://arxiv.org/abs/hep-ph/9509348}{{\tt
  hep-ph/9509348}}].

\bibitem{Kovner:1999bj}
A.~Kovner and J.~G. Milhano, \emph{{Vector potential versus color charge
  density in low x evolution}},
  \href{http://dx.doi.org/10.1103/PhysRevD.61.014012}{\emph{Phys. Rev.} {\bf
  D61} (2000) 014012}, [\href{https://arxiv.org/abs/hep-ph/9904420}{{\tt
  hep-ph/9904420}}].

\bibitem{Iancu:2000hn}
E.~Iancu, A.~Leonidov and L.~D. McLerran, \emph{{Nonlinear gluon evolution in
  the color glass condensate. 1.}},
  \href{http://dx.doi.org/10.1016/S0375-9474(01)00642-X}{\emph{Nucl. Phys.}
  {\bf A692} (2001) 583--645},
  [\href{https://arxiv.org/abs/hep-ph/0011241}{{\tt hep-ph/0011241}}].

\bibitem{Dumitru:2005gt}
A.~Dumitru, A.~Hayashigaki and J.~Jalilian-Marian, \emph{{The Color glass
  condensate and hadron production in the forward region}},
  \href{http://dx.doi.org/10.1016/j.nuclphysa.2005.11.014}{\emph{Nucl. Phys.}
  {\bf A765} (2006) 464--482},
  [\href{https://arxiv.org/abs/hep-ph/0506308}{{\tt hep-ph/0506308}}].

\bibitem{Deak:2009xt}
M.~Deak, F.~Hautmann, H.~Jung and K.~Kutak, \emph{{Forward Jet Production at
  the Large Hadron Collider}},
  \href{http://dx.doi.org/10.1088/1126-6708/2009/09/121}{\emph{JHEP} {\bf 09}
  (2009) 121}, [\href{https://arxiv.org/abs/0908.0538}{{\tt 0908.0538}}].

\bibitem{vanHameren:2013fla}
A.~van Hameren, P.~Kotko and K.~Kutak, \emph{{Three jet production and gluon
  saturation effects in p-p and p-Pb collisions within high-energy
  factorization}}, \href{http://dx.doi.org/10.1103/PhysRevD.88.094001,
  10.1103/PhysRevD.90.039901}{\emph{Phys. Rev.} {\bf D88} (2013) 094001},
  [\href{https://arxiv.org/abs/1308.0452}{{\tt 1308.0452}}].

\bibitem{vanHameren:2016kkz}
A.~van Hameren, \emph{{KaTie : For parton-level event generation with
  $k_T$-dependent initial states}},
  \href{http://dx.doi.org/10.1016/j.cpc.2017.11.005}{\emph{Comput. Phys.
  Commun.} {\bf 224} (2018) 371--380},
  [\href{https://arxiv.org/abs/1611.00680}{{\tt 1611.00680}}].

\bibitem{Jung:2010si}
H.~Jung et~al., \emph{{The CCFM Monte Carlo generator CASCADE version 2.2.03}},
  \href{http://dx.doi.org/10.1140/epjc/s10052-010-1507-z}{\emph{Eur. Phys. J.}
  {\bf C70} (2010) 1237--1249}, [\href{https://arxiv.org/abs/1008.0152}{{\tt
  1008.0152}}].

\bibitem{Cacciari:2008gp}
M.~Cacciari, G.~P. Salam and G.~Soyez, \emph{{The anti-$k_t$ jet clustering
  algorithm}},
  \href{http://dx.doi.org/10.1088/1126-6708/2008/04/063}{\emph{JHEP} {\bf 04}
  (2008) 063}, [\href{https://arxiv.org/abs/0802.1189}{{\tt 0802.1189}}].

\bibitem{Buckley:2014ana}
A.~Buckley, J.~Ferrando, S.~Lloyd, K.~Nordström, B.~Page, M.~Rüfenacht
  et~al., \emph{{LHAPDF6: parton density access in the LHC precision era}},
  \href{http://dx.doi.org/10.1140/epjc/s10052-015-3318-8}{\emph{Eur. Phys. J.}
  {\bf C75} (2015) 132}, [\href{https://arxiv.org/abs/1412.7420}{{\tt
  1412.7420}}].

\bibitem{Hautmann:2019biw}
F.~Hautmann, L.~Keersmaekers, A.~Lelek and A.~M. Van~Kampen, \emph{{Dynamical
  resolution scale in transverse momentum distributions at the LHC}},
  \href{http://dx.doi.org/10.1016/j.nuclphysb.2019.114795}{\emph{Nucl. Phys.}
  {\bf B949} (2019) 114795}, [\href{https://arxiv.org/abs/1908.08524}{{\tt
  1908.08524}}].

\bibitem{Nefedov:2020ecb}
M.~A. Nefedov, \emph{{Parton Reggeization Approach for gluon-induced processes
  at Next-to-Leading order}},  \href{https://arxiv.org/abs/2003.02194}{{\tt
  2003.02194}}.

\bibitem{Golec-Biernat:2018hqo}
K.~Golec-Biernat and A.~M. Stasto, \emph{{On the use of the KMR unintegrated
  parton distribution functions}},
  \href{http://dx.doi.org/10.1016/j.physletb.2018.04.061}{\emph{Phys. Lett.}
  {\bf B781} (2018) 633--638}, [\href{https://arxiv.org/abs/1803.06246}{{\tt
  1803.06246}}].

\bibitem{Guiot:2019vsm}
B.~Guiot, \emph{{Pathologies of the Kimber-Martin-Ryskin prescriptions for
  unintegrated PDFs: Which prescription should be preferred?}},
  \href{http://dx.doi.org/10.1103/PhysRevD.101.054006}{\emph{Phys. Rev.} {\bf
  D101} (2020) 054006}, [\href{https://arxiv.org/abs/1910.09656}{{\tt
  1910.09656}}].

\bibitem{Hautmann:2017fcj}
F.~Hautmann, H.~Jung, A.~Lelek, V.~Radescu and R.~Zlebcik, \emph{{Collinear and
  TMD Quark and Gluon Densities from Parton Branching Solution of QCD Evolution
  Equations}}, \href{http://dx.doi.org/10.1007/JHEP01(2018)070}{\emph{JHEP}
  {\bf 01} (2018) 070}, [\href{https://arxiv.org/abs/1708.03279}{{\tt
  1708.03279}}].

\bibitem{Martinez:2018jxt}
A.~Bermudez~Martinez, P.~Connor, H.~Jung, A.~Lelek, R.~Žlebčík, F.~Hautmann
  et~al., \emph{{Collinear and TMD parton densities from fits to precision DIS
  measurements in the parton branching method}},
  \href{http://dx.doi.org/10.1103/PhysRevD.99.074008}{\emph{Phys. Rev.} {\bf
  D99} (2019) 074008}, [\href{https://arxiv.org/abs/1804.11152}{{\tt
  1804.11152}}].

\bibitem{Hautmann:2014kza}
F.~Hautmann, H.~Jung, M.~Krämer, P.~J. Mulders, E.~R. Nocera, T.~C. Rogers
  et~al., \emph{{TMDlib and TMDplotter: library and plotting tools for
  transverse-momentum-dependent parton distributions}},
  \href{http://dx.doi.org/10.1140/epjc/s10052-014-3220-9}{\emph{Eur. Phys. J.}
  {\bf C74} (2014) 3220}, [\href{https://arxiv.org/abs/1408.3015}{{\tt
  1408.3015}}].

\bibitem{Kutak:2014wga}
K.~Kutak, \emph{{Hard scale dependent gluon density, saturation and
  forward-forward dijet production at the LHC}},
  \href{http://dx.doi.org/10.1103/PhysRevD.91.034021}{\emph{Phys. Rev.} {\bf
  D91} (2015) 034021}, [\href{https://arxiv.org/abs/1409.3822}{{\tt
  1409.3822}}].

\bibitem{vanHameren:2019ysa}
A.~van Hameren, P.~Kotko, K.~Kutak and S.~Sapeta, \emph{{Broadening and
  saturation effects in dijet azimuthal correlations in p-p and p-Pb collisions
  at $\mathbf{\sqrt{s}} = $ 5.02 TeV}},
  \href{http://dx.doi.org/10.1016/j.physletb.2019.06.055}{\emph{Phys. Lett.}
  {\bf B795} (2019) 511--515}, [\href{https://arxiv.org/abs/1903.01361}{{\tt
  1903.01361}}].

\bibitem{Marquet:2019ltn}
C.~Marquet, S.-Y. Wei and B.-W. Xiao, \emph{{Probing parton saturation with
  forward $Z^0$-boson production at small transverse momentum in p+p and p+A
  collisions}},
  \href{http://dx.doi.org/10.1016/j.physletb.2020.135253}{\emph{Phys. Lett.}
  {\bf B802} (2020) 135253}, [\href{https://arxiv.org/abs/1909.08572}{{\tt
  1909.08572}}].

\bibitem{Bury:2016cue}
M.~Bury, M.~Deak, K.~Kutak and S.~Sapeta, \emph{{Single and double inclusive
  forward jet production at the LHC at $\sqrt{s}$ = 7 and 13 TeV}},
  \href{http://dx.doi.org/10.1016/j.physletb.2016.07.041}{\emph{Phys. Lett. B}
  {\bf 760} (2016) 594--601}, [\href{https://arxiv.org/abs/1604.01305}{{\tt
  1604.01305}}].

\bibitem{Kotko:2016lej}
P.~Kotko, A.~Stasto and M.~Strikman, \emph{{Exploring minijets beyond the
  leading power}},
  \href{http://dx.doi.org/10.1103/PhysRevD.95.054009}{\emph{Phys. Rev. D} {\bf
  95} (2017) 054009}, [\href{https://arxiv.org/abs/1608.00523}{{\tt
  1608.00523}}].

\bibitem{Blanco:2019qbm}
E.~Blanco, A.~van Hameren, H.~Jung, A.~Kusina and K.~Kutak, \emph{{$Z$ boson
  production in proton-lead collisions at the LHC accounting for transverse
  momenta of initial partons}},
  \href{http://dx.doi.org/10.1103/PhysRevD.100.054023}{\emph{Phys. Rev. D} {\bf
  100} (2019) 054023}, [\href{https://arxiv.org/abs/1905.07331}{{\tt
  1905.07331}}].

\end{thebibliography}\endgroup

\end{document}